\documentclass[11pt,letterpaper]{article}

\usepackage{amsmath}
\usepackage{amsthm}
\usepackage{amssymb}
\usepackage{mathtools}
\usepackage{dsfont}
\usepackage{authblk}
\usepackage[dvipsnames]{xcolor}

\newcommand{\comment}[1]{}
\newcommand{\Omit}[1]{}

\usepackage{color}

\newtheorem{theorem}{Theorem}[section]
\newtheorem{lemma}[theorem]{Lemma}

\newtheorem{claim}[theorem]{Claim}

\newcommand{\expect}[2]{\mathop{\mathbb{E}}_{#1} \left[ #2 \right]}

\newcommand{\price}{\bf{p}}

\newcommand{\Left}{L}
\newcommand{\Right}{R}

\begin{document}

\title{A note on the Economic Interpretation of Online Bipartite Matching}

\title{An Economics-Based Analysis of RANKING for Online Bipartite Matching}

\author[1]{Alon Eden}
\author[2]{Michal Feldman} 
\author[2]{Amos Fiat}
\author[2]{Kineret Segal}
\affil[1]{Harvard University}
\affil[2]{Tel Aviv University}
\date{}

\maketitle

\begin{abstract}
In their seminal paper, Karp, Vazirani and Vazirani (STOC'90) introduce the online bipartite matching problem, and the RANKING algorithm, which admits a tight $1-\frac{1}{e}$ competitive ratio.
Since its publication, the problem has received considerable attention, including a sequence of simplified proofs. In this paper we present a new proof that gives an economic interpretation of the RANKING algorithm --- further simplifying the proof and avoiding arguments such as duality.
The new proof gives a new perspective on previous proofs.

\end{abstract}


\section{The Online Bipartite Matching Problem}

Consider a bipartite graph $G= (\Left,\Right; E)$, where $\Left$ and $\Right$ are the sets of left and right vertices, respectively, and $E \subseteq L \times R$ is the set of edges.
A matching in $G$ is a subset of the  $M \subseteq E$ such that no vertex is incident to more than one edge in $M$.

In online bipartite matching problem, introduced in the seminal work of Karp, Vazirani and Vazirani \cite{DBLP:conf/stoc/KarpVV90}, the set of edges, $E$, is initially unknown. The vertices in $R$ are all present at the start, but vertices from $L$, along with their edges, appear over time. The $i$'th vertex, $\ell_i \in L$, along with all edges from $\ell_i$ to vertices in $R$ appear (simultaneously) at the $i$th time step. Immediately and irrevocably, the online algorithm must decide which, if any, vertex $r\in R$ is to be matched to the new vertex $\ell_i\in L$.

The objective of the online algorithm is to maximize the cardinality of the final matching (after all vertices in $L$ have appeared).

The online bipartite matching problem and variants thereof have received a surge of interest in recent years due to their connection to allocation problems in economic settings, such as internet advertising applications. The new problems are theoretically appealing and practically relevant. See Mehta \cite{Mehta13} and references therein for a survey.

For ease of exposition, we assume that $\Left=\{\ell_1, \ldots, \ell_n\}$, $\Right=\{\ell_1, \ldots, \ell_n\}$, and $G$ admits a matching of size $n$.
However, the results and proofs in this paper hold for arbitrary bipartite graphs, where $|\Left|$ and $|\Right|$ may be arbitrary and $n$ is the size of the maximum matching ($n$ is unknown apriori).

A family of simple greedy algorithms for the online bipartite matching problem match every arriving vertex with an arbitrary unmatched neighbor, if available. (This is a family of algorithms as the arbitrary choices may be different).
Every such greedy algorithm outputs a maximal matching, hence has cardinality of at least $n/2$.
It is easy to see that this bound is tight; {\sl i.e.}, there exist graphs for which this greedy algorithm cannot achieve more than half of the maximum matching (for example, the first $n/2$ vertices in $L$ are connected to all vertices in $\Right$, and the remaining $n/2$ vertices in $\Left$ are connected to the $n/2$ vertices in $\Right$ that were matched to the first half of the vertices in $\Left$.)
A randomized version of the greedy algorithm, which chooses a currently unmatched neighbor uniformly at random (if one exists), cannot improve upon this ratio either (up to lower order terms) \cite{DBLP:conf/stoc/KarpVV90}.

\paragraph{The Randomized RANKING Algorithm.}
The RANKING algorithm, introduced by Karp, Vazirani and Vazirani \cite{DBLP:conf/stoc/KarpVV90}, is a simple randomized algorithm, which works as follows: it first chooses a random permutation $\pi$ over the vertices in $\Right$.
Upon the arrival of a vertex $\ell_i$, RANKING matches $\ell_i$ to the highest-ranked (with respect to $\pi$) currently unmatched neighbor of $\ell_i$.

Karp et al. proved that RANKING matches at least $(1-\frac{1}{e})n$ edges in expectation. They also showed that this bound is tight (up to low order terms{\footnote{That is, no online algorithm matches more than $(1-1/e)n+O(1)$ edges in expectation. Specifically, Feige \cite{Feige19} shows that no online integral matching algorithm matches more than $(1-\frac{1}{e})n+1-\frac{2}{e}+O(\frac{1}{n!})$ edges in expectation.}).

The analysis in the original paper was quite complicated (and imprecise in places).
Subsequent papers by Goel and Mehta \cite{DBLP:conf/soda/GoelM08}, Birnbaum and
Mathieu \cite{DBLP:journals/sigact/BirnbaumM08} and Devanur, Jain and Kleinberg \cite{DBLP:conf/soda/DevanurJK13} simplified the analysis considerably.
In this work we present an arguably even simpler proof, which is based on an economic interpretation of the online bipartite matching problem.
It bears similarities to the proof of \cite{DBLP:conf/soda/DevanurJK13}, but does not make an explicit use of linear programming duality.


\section{An Economics-Based Analysis of RANKING}

In what follows we give an economic interpretation of the online bipartite matching.
Given a graph $G = (\Left,\Right; E)$, vertices of $\Right$ represent items, and vertices of $\Left$ represent utility maximizing {\em unit-demand} buyers. 

For every vertex $x \in \Left \cup \Right$, let $N(x)$ denote the neighbors of $x$, {\sl e.g.,}, $N(\ell_i)=\{r_j \mid (\ell_i,r_j)\in E\}$.
Let $v_i(r_j)$ denote the value of buyer $\ell_i$ for an item $r_j$.
For our purposes assume that $v_i(r_j)=1$  if $r_j\in N(\ell_i)$, and $v_i(r_j)=0$ otherwise. For a set of items $X\subset R$, let $v_i(X)=\max_{\{r\in X\}}v_i(r).$


Consider the following process, hereafter referred to as the {\em market process}:
Before the arrival of any buyer, every item $r_j$ is assigned a price $p_j$.
The {\em utility} a buyer $\ell_i$ derives from an item $r_j$ is
$$
u_i(r_j) = v_i(r_j)-p_j.
$$
Buyers arrive online, in arbitrary order.
Upon arrival, a buyer chooses an item that maximizes her utility amongst all remaining items (possibly choosing no item).
This process induces a matching $M$, where $(\ell_i,r_j)\in M$ if buyer $\ell_i$ chooses item $r_j$.
The {\em social welfare} of a matching $M$ is the sum of buyer valuations for their items, i.e.,
$$SW(M)=\sum_{i\in [n]}v_i(M(i)),$$
where $M(i)$ denotes the item chosen by buyer $\ell_i$ (possibly an empty set).
Since every buyer that receives an item has value 1, the social welfare of a matching $M$ equals the cardinality of $M$.

The following claim shows a connection between the  market process and the randomized RANKING algorithm.

\begin{claim}
\label{obs:equivalent}
Consider the market process above.
Let $D$ be some arbitrary distribution
over $[a,b]$, $0 \leq a < b\leq 1$,
with no point mass,
and choose the price of item $r_j$, $p_j$, independently from $D$, $p_j \thicksim D.$
Then, the resulting distribution over matchings is identical to the one obtained by the randomized RANKING algorithm.
\end{claim}



\begin{proof}
Recall that RANKING chooses a random permutation $\pi$ over $\Right$ and upon the arrival of a vertex $\ell_i \in \Left$, $\ell_i$ is matched to the highest-ranked (according to $\pi$) available vertex in $\Right$.

Setting a random price, $p_j$, for items $r_j$ implies that the utility of item $r_j$ is also random ($1-p_j$) --- given that there is an edge from the buyer to the item. Buyers always choose the maximal utility item. Ergo,
a random permutation $\pi$ over $\Right$ is equivalent to random prices, chosen from $D$, to items $r_j\in R$.

The range $[a,b]$, $b\leq 1$, is to avoid negative utility and no point mass implies no equality in pricing or utility.
\end{proof}

We now use Claim~\ref{obs:equivalent} to
prove that RANKING matches at least $(1-\frac{1}{e})n$ edges in expectation. It suffices to show some distribution $D$, and prices independently chosen from $D$, so that the expected social welfare in the market process is at least $(1-\frac{1}{e})n$.

A particular distribution of interest is the distribution $\widehat{D}$ from which one samples as follows: choose a value $w$ uniformly at random in the interval $[0,1]$ and return  $e^{w-1}$.

\begin{theorem}
\label{thm:welfare}
The market process where prices are sampled independently from $\widehat{D}$, $p_j\sim \widehat{D}$, for $j=1,\ldots,n$, gives an expected social welfare of at least $(1-\frac{1}{e})n$.
\end{theorem}

\begin{proof}
To prove the theorem, we decompose the social welfare into the sum of buyer utilities and seller's revenue (this technique has proved useful in other settings \cite{FeldmanGL15,DuettingFKL17,EhsaniHKS18}).
For every item $r_j$, let $\mathrm{rev}_j$ denote the revenue obtained by $r_j$ (i.e., $p_j$ if the item was purchased and 0 otherwise).
For every buyer $\ell_i$, let $\mathrm{util}_i$ denote the utility of buyer $\ell_i$; i.e.,
$$
\mathrm{util}_i=
\begin{cases}
1-p_j, & \text{if buyer } \ell_i \text{ purchases item } r_j \in N(\ell_i)\\
0, & \text{if buyer } \ell_i \text{ does not purchase any item.}\\
\end{cases}
$$

Fix some (arbitrary) arrival order of the buyers and a price vector ${\price}=(p_1, \ldots, p_n)$, and
let $M$ be the matching resulting from this process.
The following equation gives the resulting social welfare as the sum of the buyer utilities and the total revenue:
\begin{equation}
\sum \limits _{\ell_i\in \Left} \mathrm{util}_i + \sum \limits _{r_j \in \Right} \mathrm{rev}_j =
\sum\limits_{(\ell_i,r_j)\in M} (1-p_j)  +  p_j = |M|.\label{eq:utilrev}
\end{equation}

We now give the key lemma used in the proof of Theorem~\ref{thm:welfare}.
\begin{lemma}
\label{cl:sufficient}
Assume prices are sampled independently from $\widehat{D}$, $p_j\sim \widehat{D}$. Then, for every edge $(\ell_i,r_j) \in E$,
$$\expect{\bf{w}}{\mathrm{util}_i + \mathrm{rev}_j} \geq 1-\frac{1}{e}.$$
\end{lemma}

Before proving Lemma~\ref{cl:sufficient}, we show that it implies Theorem~\ref{thm:welfare}.
Fix a maximum matching $M^*$ and let $M$ be the matching resulting from the market process above.
Using Equation (\ref{eq:utilrev}), linearity of expectation, and Claim~\ref{cl:sufficient} we get that:
\begin{eqnarray*}
\expect{\bf{w}}{|M|} & = &\expect{\bf{w}} {\sum\limits_i \mathrm{util}_i+ \sum\limits_j \mathrm{rev}_j} \geq
\expect{\bf{w}} {\sum\limits_{(\ell_i,r_j) \in M^*} \left( \mathrm{util}_i + \mathrm{rev}_j \right)}\\
&= &\sum\limits_{(\ell_i,r_j) \in M^*} \expect{\bf{w}} {\mathrm{util}_i + \mathrm{rev}_j}
\geq \left(1-\frac{1}{e}\right)|M^*|.
\end{eqnarray*}

We conclude the proof  of Theorem \ref{thm:welfare} by proving Lemma \ref{cl:sufficient}.
\begin{proof}[Proof of Lemma \ref{cl:sufficient}]
	Fix some arbitrary order of buyer arrival $\sigma$, buyer $\ell_i$ and item $r_j$ such that $(\ell_i,r_j)\in E$. Sample item prices from $\widehat{D}$. Let $M_{-j}$ denote the matching produced in a market without item $r_j$ (under the same arrival order $\sigma$).
	Let $p=e^{y-1}$ be the price of the item matched to buyer $\ell_i$ in $M_{-j}$; if $\ell_i$ is left unmatched, set $p=1$ (in this case $p=e^{y-1}$ for $y=1$).
We make the following two simple observations, essentially trivial when thinking in terms of buyers and sellers:
	\begin{itemize}
		\item {\bf Observation (1):} If $p_j < p$, then item $r_j$ is sold. This follows since buyer $\ell_i$ derives higher utility from item $r_j$ than from its match in $M_{-j}$, so either item $r_j$ is purchased by an earlier buyer, or buyer $\ell_i$ buys it upon arrival (note that in the case where $p=1$ and $r_j$ is unsold upon $\ell_i$'s arrival, buyer $\ell_i$ gains by purchasing item $r_j$ for a price $p_j < 1$).
		\item {\bf Observation (2):} It holds that $\mathrm{util}_i \geq 1-p$. This observation follows since $1-p$ is the utility of buyer $\ell_i$ in $M_{-j}$, and the introduction of an additional item into the market (item $r_j$ in our case) can never decrease the utility of any buyer. This last claim holds since, by induction, every buyer faces the same set of items available to her plus, possibly, one additional item. This is obviously true for the first incoming buyer, and remains true subsequently since the introduction of an additional item never induces a buyer to purchase an item previously waived.
		
	\end{itemize}
	
	Let $y \in [0,1]$ be the value such that $p=e^{y-1}$. It follows that
	$$\expect{\bf{w}}{\mathrm{rev}_j}  =  \expect{\bf{w}}{p_j\cdot \mathds{1}\left[r_j \mbox{ is sold}\right]}
	\geq \expect{\bf{w}}{p_j\cdot \mathds{1}\left[p_j<p\right]} = \int_0^{y}e^{x-1}dx  = e^{y-1}-\frac{1}{e}=p-\frac{1}{e},$$
	where the inequality follows by observation (1).
	It now follows from observation (2) that	
	$$\expect{\bf{w}}{\mathrm{util}_i + \mathrm{rev}_j}  \geq  1-p +p-\frac{1}{e}=1-\frac{1}{e},$$
as desired.
	
\end{proof}
This concludes the proof of Theorem~\ref{thm:welfare}.
\end{proof}



Note that our proof is closely related to the proof presented by Devanur et al. \cite{DBLP:conf/soda/DevanurJK13} (and further simplifications by Mathieu \cite{Mathieu2011}), which is based on linear programming duality. Indeed, the utility and revenue terms used in our analysis are essentially scaled versions of the dual variables in \cite{DBLP:conf/soda/DevanurJK13}. However, the economic interpretation introduced in this paper simplifies the proof even further. In particular, it does not make an explicit use of linear programming duality, and thus eliminates the need to argue about the dual program and its feasibility altogether. Some of the other arguments in \cite{DBLP:conf/soda/DevanurJK13} are more readily apparent when viewed from the economic perspective.


%
%

\paragraph{Acknowledgements.}
This project has received funding from the European Research Council (ERC) under the European Union's Horizon 2020 research and innovation program (grant agreement No. 866132), and by the Israel Science Foundation (grant number 317/17).

\bibliographystyle{unsrt}
\bibliography{onlineMatching}

\end{document}